\documentstyle[12pt, amsfonts, amssymb]{article}

\date{\today}

\begin{document}

\def\be{\begin{equation}}
\def\ee{\end{equation}}
\def\ba{\begin{eqnarray}}
\def\ea{\end{eqnarray}}
\def\lb{\label}

\def\a{\alpha}
\def\b{\beta}
\def\g{\gamma}
\def\d{\delta}
\def\i{\eta}
\def\e{\varepsilon}
\def\l{\lambda}
\def\s{\sigma}
\def\t{\tau}
\def\r{\rho}
\def\v{\varphi}

\def\D{\Delta}
\def\G{\Gamma}
\def\L{\Lambda}
\def\P{\Phi}

\def\E{{\cal E}}

\def\p{\hat p}
\def\fp{{\frak p}}

\def\C{\Bbb C}
\def\R{\Bbb R}
\def\Q{\Bbb Q}
\def\Z{\Bbb Z}
\def\F{\Bbb F}

\def\uz{\underline z}

\def\Vp{{\cal V}_p}
\def\Hp{{\cal H}_p}

\def\bq{\overline{q}}

\def\id{\mbox{\rm 1\hspace{-3.5pt}I}}
\def\1{1\!\!{\rm I}}

\def\eod{\phantom{a}\hfill \rule{2.5mm}{2.5mm}}

\def\hF{\hat{F}}
\def\hA{\hat{A}}
\def\hB{\hat{B}}

\def\R{\hat{R}}
\def\Rp{\hat{R}(p)}

\def\uz{\underline z}
\def\pz{\Pi_{23}\;\uz}

\hyphenation{ho-lo-no-my}


\hfill{CERN-TH/2000-362}

\vspace{1cm}

\begin{center}

{\huge Monodromy representations\\ \vspace{3mm} of the braid group}
\footnote{ Plenary talk, presented by I.T. Todorov 
at the XXIII International Colloquium on Group Theoretical Methods 
in Physics, Dubna, July 31 -- August 5, 2000} 

\vspace{1cm}

{\large I.T. Todorov} \footnote{E-mail: todorov@inrne.bas.bg,
Ivan.Todorov@cern.ch, itodorov@esi.ac.at}$^{,4}$\,
and {\large L.K. Hadjiivanov} \footnote{E-mail: lhadji@inrne.bas.bg,
Ludmil.Hadjiivanov@cern.ch}$^{,}$\footnote{Permanent
address: Theoretical Physics Division,  
Institute for Nuclear Research and Nuclear Energy, 
Tsarigradsko Chaussee 72, BG-1784 Sofia, Bulgaria }

\vspace{3mm}
{\footnotesize 
Theory Division, CERN, CH-1211 Geneva 23} 

\end{center}

\vspace{1cm}

\begin{abstract}

Chiral conformal blocks in a rational conformal field theory are a
far going extension of Gauss hypergeometric functions. The
associated monodromy representations of Artin's braid group
${\cal B}_n\,$ capture the essence of the modern view on the 
subject, which originates in ideas of Riemann and Schwarz.

Physically, such monodromy representations correspond to a new type of
braid group statistics, which may manifest itself in two-dimensional 
critical phenomena, e.g. in some exotic quantum Hall states. The
associated primary fields satisfy $R$-matrix exchange relations.
The description of the internal symmetry of such fields requires
an extension of the concept of a group, thus giving room to
quantum groups and their generalizations.

We review the appearance of braid group representations in the space of
solutions of the Knizhnik--Zamolodchikov equation, with an emphasis on the
role of a regular basis of solutions which allows us to treat 
the case of indecomposable representations of ${\cal B}_n\,$ as well.

\end{abstract}

\vfill

\section*{Introduction}

Artin's braid group ${\cal B}_n\,$ -- with its monodromy
representations -- is a good example of a focal point for
important developments in both mathematics and physics.

In {\em mathematics} it appears in the description of topological
invariants of algebraic functions \cite{Ar68} and the related study of
multiparametric integrals and (generalized) hypergeometric
functions [2 -- 4]  
as well as in the theory of knot invariants and
invariants of three-dimensional manifolds [5 -- 9]. 
The main {\em physical applications} go under
the heading of {\em generalized statistics} (anticipated by Arnold
in \cite{Ar68} -- see Section 1). The {\em
Knizhnik--Zamolodchikov} (KZ) {\em equation} (Section 2) is a
common playground for physicists and mathematicians.

We illustrate highbrow mathematical results of \cite{Ko87} and
\cite{Dr89}
on the relation between the monodromy representations of ${\cal
B}_n\,$ in the space of solutions of the KZ equation, for a
semisimple Lie algebra ${\cal G}\,,$ and the universal $R$-matrix
for $U_q({\cal G})\,$ by simple computations for the special case
of ${\cal G} = su(N)\,$ step operators and ${\cal B}_3\,$ (Section 3).
In fact, we go in our explicit construction beyond these general results
by also treating on an equal footing the indecomposable representations of
${\cal B}_3\,$ for $q\,$ an even root of unity $( q^h = -1\, )\,.$

\section{Permutation and braid group statistics}
\setcounter{equation}{0}
\renewcommand{\theequation}{1.\arabic{equation}}

The symmetry of a one-component wave function $\Psi (x_1,\dots ,
x_n)\,$ is described by either of the one-dimensional
representations of the group ${\cal S}_n\,$ of permutations giving rise
to Bose and/or Fermi statistics. Multicomponent wave functions
corresponding to particles with internal quantum numbers, may
transform under higher dimensional representations of ${\cal S}_n\,$
corresponding to parastatistics. If one allows for multivalued
wave functions, then the exchange of two arguments $x\,$ and $y\,$
may depend on the (homotopy type of the) path along which $x\,$
and $y\,$ are exchanged, giving rise to a representation of
the {\em braid group} ${\cal B}_n\,$ of {\em $n\,$ strands}.

${\cal B}_n$ was defined by E. Artin in 1925 as a group with
$n-1\,$ generators, $b_1 ,\dots , b_{n-1}\,$ ($b_i\,$ ``braiding"
the strands $i\,$ and $i+1\,$), satisfying the following two
relations:
\begin{equation}
b_i\, b_j\, =\, b_j\, b_i\,,\ |i-j|\ge 2\,,\quad
b_i\, b_{i+1}\, b_i\, = \, b_{i+1}\, b_i\, b_{i+1}\,,\
i=1,\dots , n-2\,.
\label{1.1}
\end{equation}
Let $\s : {\cal B}_n\ \rightarrow \ {\cal S}_n\,$ be the group
homomorphism defined by
\begin{equation}
\s (b_i )\, =\, \s_i\,,\quad {\s_i}^2\,=\, 1\, ( \in {\cal
S}_n\,)\,,\qquad i=1,\dots ,n-1\,,
\label{1.2}
\end{equation}
where $\s_i\,$ are the basic transpositions exchanging $i\,$ and
$i+1\,$ that generate ${\cal S}_n\,.$ The kernel ${\cal P}_n\,$ of
this homomorphism is called the {\em monodromy} (or pure braid)
{\em group}. Note that the element
\begin{equation}
c^n\, :=\, (b_1 b_2 \dots b_{n-1} )^n
\label{1.3}
\end{equation}
generates the centre of ${\cal B}_n\,.$

The braid group ${\cal B}_n\,$ and its invariant subgroup
${\cal P}_n\,$ have a topological interpretation. Consider the
$n$-dimensional manifold
\begin{equation}
\lb{1.4}
Y_n = {\C}^n \setminus {\rm Diag} \equiv \{ {\vec z} = (z_1 ,\dots ,
z_n\, ) \in {\C}^n\,;\ i\ne j\,\Rightarrow\, z_i\ne z_j\,\}
\end{equation}
($Y_n\,$ is the analyticity domain of $n$-point conformal blocks in chiral
conformal field theory). The symmetric group ${\cal S}_n\,$ acts on
$Y_n\,$ by permutations of coordinates. The factor space $X_n =
Y_n / {\cal S}_n\,$ is the {\em configuration space of $n$ points}
(``identical particles") in ${\C}^n\,.$

\vspace{4mm}

\noindent
{\bf Proposition 1.1~} (\cite{Ar68})\ {\em The braid group ${\cal
B}_n\,$ is isomorphic to the fundamental group
$\pi_1 (X_n , {\vec z}_0 )\,$ of the configuration space {\rm (}for, say,
${\vec z}_0\, = (n\,,\dots ,1)\,${\rm );} similarly,
${\cal P}_n \simeq\pi_1 (Y_n , {\vec z}_0 )\,.$ }

Clearly, had we substituted the complex plane, ${\C} \simeq
{\mathbb R}^2\,,$ by an $s$-dimensional space ${\R}^s\,$ for
$s\ge 3\,,$ the fundamental group
$\pi_1 ( ({\R}^s )^{\otimes n} \setminus {\rm Diag}\,, {\vec z}_0 )\,$
would have been trivial, and no interesting relation with the braid
group could be expected. This simple topological fact explains why
the possibilities for generalized statistics are richer in low
(one- and two-) dimensional physics. One may wonder why it took 
more than half a century after the appearance of Bose and
Fermi statistics in quantum mechanics for such a basic
observation to gradually find its way into the physics literature.
The notion of a braid group seems to have first appeared in physics in
connection with the Dirac string \cite{Newman42}; the peculiarity of low
dimensional statistics was noted in \cite{FinkRub68}; the problem
was then treated more systematically in the framework of quantum mechanics 
\cite{LM77} 
and in the context of local current
algebras \cite{GMS80}. Moreover, this pioneering work did not attract much
attention before  it was repeated by others (starting with
\cite{Wil82}) when catch-words such as ``anyons" were coined.
The story of the ancestry of the ``anyon" has once been told
with authority \cite{BLSW}, but it continues to be ignored in ``mainstream 
publications".

A deeper understanding of particle statistics came from the
``algebraic'' study of superselection sectors in local quantum field
theory (see \cite{Ha92} and \cite{DR90} where earlier work of Doplicher,
Haag and Roberts is also cited). We offer here an informal
(physicist's oriented) formulation of the main result of this
work.

The starting point of the {\em algebraic} (Haag--Kastler) {\em
approach} is the concept of an {\em algebra ${\cal A}\,$ of local
observables}. 
As lucidly explained in \cite{Buch}, it provides an intrinsic, 
coordinate-free description of the algebra generated by local gauge invariant
(Wightman) Bose fields -- such as the stress-energy tensor and conserved
$U(1)\,$ currents. 
Departing slightly from the purist's algebraic view, we shall identify
from the outset ${\cal A}\,$ with its {\em vacuum representation} in a
Hilbert space ${\cal H}\,$ that carries a unitary positive energy
representation of the Poincar\'e group with a {\em unique}
translation invariant {\em vacuum state}. It is important
that gauge-dependent charge carrying (and/or multivalued) fields
are excluded from ${\cal A}\,.$ They reappear -- as derived
objects -- in the role of intertwiners among inequivalent
representations of ${\cal A}\,.$

{\em Superselection sectors} are defined by {\em irreducible
positive energy representations} of ${\cal A}\,$ that can be
obtained from the {\em vacuum sector} by the action of {\em
localizable ``charged fields"} -- i.e. of fields that commute at
space-like separations with the observables (but need not be local
among themselves). Products of charged fields acting on the vacuum
give rise, typically, to a finite sum of superselection sectors
defining the {\em fusion rules} of the theory. (To make this
precise one needs, in fact, more elaborate tools -- such as
*-endomorphisms of a completion of ${\cal A}\,$ that are
localizable in space-like cones; the shortcoming of a simple-minded use
of ``charged fields" is the non-uniqueness of their choice and
hence, of the multiplicities entering the above naive definition of
fusion rules.) A fancy way to express the fact that there is a
well defined composition law for representations of ${\cal A}\,$
(analogous to tensor product of group representations) is to say
that {\em superselection sectors give rise to a tensor category}.
A memorable result of Doplicher and Roberts \cite{DR90} crowning two
decades of imaginative work of Haag's school 
says that, for a local quantum theory with no massless
excitations in space-time dimensions
$D\ge 4\,,$ this category is equivalent to the category of
irreducible representations (IR) of a compact group $G\,.$ In more
down-to-earth terms it means that $G\,$ acts -- by automorphisms
-- on charged fields as a {\em gauge group of the first kind}
(recall that a gauge group leaves all observables invariant, not
only the Hamiltonian). Superselection sectors are labelled by
(equivalence classes of) IR $p\in \hat G\,$ (borrowing the
terminology of representation theory of semisimple compact Lie
groups, we shall call the labels $p\,$ {\em weights}). The state
space of the theory can thus be viewed as a direct sum of tensor
product spaces:
\begin{equation}
\lb{1.5}
{\cal H}\, =\, \bigoplus_{p\in \hat G}\, {\cal H}_p \otimes {\cal
F}_p\,,\qquad d(p) :=\, {\rm dim}\, {\cal F}_p < \infty\,,
\end{equation}
where ${\cal F}_p\,$ are irreducible $G$-modules. The statistics
of a sector $p$ is characterized by a {\em statistics parameter}
$\l_p = \pm \frac{1}{d(p)}\,.$ If $G\,$ is abelian (the common 
case of commuting superselection sectors labelled by the ``spin
parity" $e^{2\pi i s_p}\,,$ where $s_p\,$ is the spin, and by the
values of the electric, baryonic and leptonic charge), then $d(p) =
1\,$ for all sectors and we are faced with the familiar
{\em Bose--Fermi alternative}. If $G\,$ is non-abelian and 
$d(p) =2,3,\dots ,$ then the sector $p\,$ and its conjugate
$\bar p$ (or, in physical language, the particles of type $p\,$ and
their antiparticles $\bar p\,$) obey parastatistics.
(Unfortunately, one has no such result for quantum electrodynamics. It is,
in fact, known that the electric charge cannot be localized in a
space-like cone. Although there is no indication that, say, electrons may 
obey braid group statistics, we are unable to rule it out at present.)

These results also extend to space-time dimension $D=3\,,$
provided the superselection charges can be localized in finite
regions. In more realistic $(2+1)$-dimensional systems (like a
``quantum Hall fluid" in a strong magnetic field perpendicular to
the plane of the layer), charges can be localized only in infinite
space-like cones and there is room for braid group statistics. For
$D=2\,$ braid group statistics may appear even for
(superselection) charges localized in finite domains (see
[21 - 23] and references to earlier work of these authors 
cited there). The notion of a statistics parameter
extends to this case, too, and is related to Jones index of
inclusion of associated factors of operator algebras \cite{Lon89}. It
can be written as (see \cite{FG90}, Definition 6.2)
\begin{equation}
\lb{1.6}
\l_p = \frac{1}{d(p)} e^{-2\pi i \theta_{p \bar p}}\,,\quad d(p)\,
(\, = |\l_p |^{-1}\, ) > 0\,,\quad e^{-2\pi i \theta_{p \bar p}} =
e^{- i\pi (s_p + s_{\bar p})}\,,
\end{equation}
where $s_p\,$ and $s_{\bar p}\,$ are the (fractional) spins of the
conjugate sectors $p\,$ and ${\bar p}\,.$ For $d(p) = 1\,,\ \l_p
\ne \pm 1\,,$ we are dealing with a one-dimensional representation
of the braid group, corresponding to {\em anyonic statistics}. For
non-integer $d(p)\,$ the ``gauge symmetry" of superselection sectors
cannot be described by a group. 
In fact, soon after quantum groups were introduced \cite{Drin}
the attractive possibility of applying them to describing the symmetry of
two-dimensional (conformal) models was considered by several authors
[25 -- 32] 
(refs. \cite{Fr,Reh} appearing as
predecessors of both 
[21 -- 23] and \cite{DR90}). These first
attempts did not face the problem of incompatibility between Wightman
(Hilbert space) positivity and manifest quantum group invariance. Three
different approaches have been developed to deal with this problem in
terms of: {($i$)} weak quasiHopf algebras \cite{MS92}; {($ii$)} weak
co-associative star Hopf algebras \cite{BNS99} or quantum groupoids
\cite{NV00} related recently to the Ocneanu ``double triangle algebra" and 
to boundary conformal field theory \cite{PZ}; {($iii$)} a BRS
approach with quantum group symmetry in an extended state space
\cite{DVT99,FHIOPT}. None of these developments has been fully conclusive.

To cite \cite{FG90}, ``braid statistics in two-dimensional systems is
more than a theoretical curiosity". Indeed, anyons have made their
way in the standard interpretation of the fractional Hall effect.
Non-abelian braid group statistics appears to be strongly indicated
for Hall plateaux at the second
Landau level with filling fractions
$\nu = 2 +\frac{m}{m+2}\,,\ m=2,3,\dots \,$ (cf. \cite{CGT00,FPSW00}).

\vspace{4mm}

\section{The KZ equation}
\setcounter{equation}{0}
\renewcommand{\theequation}{2.\arabic{equation}}

Let ${\cal G}\,$ be a compact Lie algebra, $V\,$ a finite-dimensional 
${\cal G}\,$ module, and $C_{ab}\,$ the (polarized) Casimir
invariant acting nontrivially on the factors $a\,$ and $b\,$ of
the $n$-fold tensor product $V^{\otimes n}\,.$ For ${\cal G} =
su(N)\,$ and $n=3\,$ we have 
\begin{equation}
\lb{2.1}
C_{12} \ (\,=\, C_{21}\, )\, =\, \left(\,
\sum_{i,j=1}^N e_{ij}\otimes e_{ji}\, -\,
\frac{1}{N}\,\sum_{i=1}^N e_{ii}\otimes\sum_{j=1}^N e_{jj}\,\right)\,
\otimes\id\,,
\end{equation}
where $e_{ij}\,$ represent the Weyl generators of $U(N)\,$ in
$V\,.$

The KZ equation is a system of partial differential equations,
which can be written compactly as
\begin{equation}
\lb{2.2}
h\, d\Psi = \sum_{1\le a<b\le n} C_{ab} \frac{d z_{ab}}{z_{ab}}
\Psi\,,\ \ z_{ab}=z_a-z_b\,,\ dz_{ab}=dz_a-dz_b\,,\ C_{ab}=C_{ba}\,;
\end{equation}
here $h\,$ is a (say, real) parameter, $\Psi = \Psi (z_1 ,\dots
,z_n)\,$ is a (regular) map, $\Psi : Y_n\rightarrow 
V^{\otimes n}\,,$ where $Y_n\,$ is ${\C}^n\,$ minus the
diagonal (see (\ref{1.4})). The system (\ref{2.2}) has a nice
geometric interpretation: it defines a {\em connection}
$\nabla = d-\Gamma\,$ on the trivial bundle $Y_n \times V^{\otimes n}\,,$ 
where $\Gamma\,$ is the connection $1$-form
\begin{equation}
\lb{2.3}
\Gamma\, =\, \frac{1}{h}\,\sum_{a<b}\, C_{ab}\,
\frac{d z_{ab}}{z_{ab}}\,.
\end{equation}
Introducing the corresponding {\em covariant derivatives}
\begin{equation}
\lb{2.4}
\nabla_a\, =\, \frac{\partial}{\partial z_a}\, -\, \frac{1}{h}\,
\sum_{b\ne a}\,\frac{C_{ab}}{z_{ab}}\,,\quad a=1,\dots ,n\,, 
\end{equation}
we can interpret (\ref{2.2}) by saying that $\Psi\,$ is
covariantly constant. This requires as a compatibility condition
the flatness of the KZ connection.

\vspace{4mm}

\noindent
{\bf Proposition 2.1~}{\em The KZ connection $\nabla =d-\Gamma\,$
has zero curvature:}
\begin{equation}
\lb{2.5}
\nabla\,\circ\,\nabla\, =\, \Gamma\,\wedge\,\Gamma\, -\,
d\,\Gamma\,=\, 0\quad\Leftrightarrow\quad [\,\nabla_a\,,\,\nabla_b\,
]\, =\, 0\,.
\end{equation}

\vspace{4mm}

The {\em proof} (see, for example, \cite{Ka95}) uses
\ba
&& [C_{ab} , C_{cd} ] = 0\quad {\rm for\ different}\ a,b,c,d\,,\\
&& [C_{ab} , C_{ac}+C_{bc} ] = 0 = [C_{ab}+C_{ac} , C_{bc}]
\quad {\rm for\ different}\ a,b,c\,,
\ea
as well as the following {\em Arnold's lemma: let}
\begin{equation}
\lb{2.8}
u_{ab} = d (\ln z_{ab} )= \frac{d z_{ab}}{z_{ab}}\
\left(\,\equiv\frac{dz_a -dz_b}{z_a - z_b}\,\right)\,;
\end{equation}
{\em then}
\begin{equation}
\lb{2.9}
u_{ab}\wedge u_{bc} + u_{bc}\wedge u_{ca} + u_{ca}\wedge u_{ab} =
0\ \ for\ \, a\ne b\ne c\ne a\,.
\end{equation}
\vspace{2mm}

The flatness of the connection $\nabla\,$ is a necessary and
sufficient condition for the holonomy group ${\cal P}_n\,$ at a
point $p\in Y_n\,$ (the transformation group in $V^{\otimes n}\,$
obtained by parallel transport of vectors along closed paths with
beginning and end in $p\,$) to give rise to a {\em (monodromy)
representation} of the fundamental group $\pi_1 (Y_n , p)\,.$

The KZ equation appears in $2D$ CFT in the context of the
Wess--Zumino--Novikov--Witten (WZNW) \cite{Wit84} model \cite{KZ84} and in
a
related study of chiral current algebras \cite{To84}. The idea of the
latter approach is simple to summarize. A {\em primary} field
$\varphi\,$ of a conformal current algebra is {\em covariant} under
two (infinite-parameter) infinitesimal transformations: under
local {\em gauge transformations} generated by the currents $J\,$
and under {\em reparametrization} generated by the stress-energy
tensor $T\,.$ On the other hand, $T\,$ is expressed quadratically
in terms of $J\,$ (by the so-called Sugawara formula). The
consistency between the two covariances and this quadratic
relation yields the {\em operator KZ equation}:
\begin{equation}
\lb{2.10}
h\, \frac{d\varphi}{d z}\, =\, : \varphi (z)\, {\vec t}\, J(z):\,.
\end{equation}
Here $C_{ab} = {\vec t}_a\otimes{\vec t}_b\,,$ the vector ${\vec
t}\,$ spanning a basis of the finite-dimensional representation of
${\cal G}\,$ such that $[ {\vec J}_0 , \varphi (z) ] = \varphi (z)\, 
{\vec t}\,$ for ${\vec J}_0 = \oint {\vec J} (z) \frac{dz}{2\pi
i}\,,$ and the ``height" $h\,$ is an integer
($h\ge N\,$ for ${\cal G} = su(N)\,$). Using also the
current-field Ward identity, we end up with Eq. (\ref{2.2}) for the
``wave function"
\begin{equation}
\lb{2.11}
\Psi (p ; z_1 ,\dots ,z_n ) = {\cal h} p |\, \varphi (z_1)\otimes
\dots \otimes \varphi (z_n) | 0 {\cal i}\,,
\end{equation}
where $p\,$ stands for the weight of the ${\cal G}$-module that
contains the bra ${\cal h} p |\,$ (see Appendix B).

The notation in Eq. (\ref{2.11}) is, in fact, ambiguous. There are,
for fixed $n\,$ and $p\,,$ several linearly independent
solutions (called {\em conformal blocks}) of the KZ equation. 
To distinguish between them one introduces the concept of a {\em
chiral vertex operator} (CVO) \cite{TK87} (the counterpart of an
intertwiner between different superselection sectors in the
algebraic approach to local quantum field theory
[19, 21 -- 23]).
We shall use instead a field $\varphi\,$ belonging to the
tensor product $V\otimes{\cal V}\,$ of a ${\cal G}\,$ and a
$U_q({\cal G})\,$ module, $\v = (\v^A_\a )\,;$ it arises naturally in
splitting the group valued field $g\,$ in the WZNW model into left
and right movers, $g^A_B(z, \bar z )= \v^A_\a (z)
({\bar\v}^{-1})^\a_B (\bar z )\,$
(see \cite{Fa90, G} and \cite{FHIOPT} for two early and a recent
paper, the latter containing some 50 more references on the
subject). Take as an example ${\cal G}=su(N)\,,\ n=3\,$ and
$\v\,$ an $SU(N)\,$ {\em step operator} (i.e. $V={\C}^N\,$
carrying the defining representation of $su(N)\,$). Then, if we
take $p\,$ to be the highest weight of the IR associated with
the Young tableau
$
\put(0,2){\framebox(5,5)}
\put(0,-3){\framebox(5,5)}
\put(5,2){\framebox(5,5)}\quad
$
with respect to both $su(N)\,$ and $U_q(sl_N)\,,$ we can reduce
(\ref{2.2}) to a system of ordinary differential equations 
for the invariant amplitude $F(\i )\,$ defined by
\begin{equation}
\lb{2.12}
\Psi (p ; z_1 , z_2 , z_3 ) \,=\, z_{13}^{-\frac{3}{4h}}\,
(\i (1-\i ))^{-\frac{N+1}{Nh}}\, F(\i )\,,\quad \i =
\frac{z_{23}}{z_{13}}\,;
\end{equation}
we find (see Appendix A):
\begin{equation}
\lb{2.13}
\left( h\frac{d}{d\i} + \frac{\Omega_{12}}{1-\i} - \frac{\Omega_{23}}{\i}
\right)\, F (\i ) =  0\,,
\end{equation}
\begin{equation}
\lb{2.14}
\Omega_{12} = C_{12} + \frac{1}{N} + 1 =
P_{12} + 1\,,\quad
\Omega_{23} = \frac{N-2}{N} - C_{12} - C_{13}\noindent
\end{equation}
($P_{12}\, (x\otimes y) = y\otimes x\, $ and $F (\i )\,$ is an
invariant $SU(N)\,$ tensor,
$F (\i ) \in {\rm Inv}\, (V^*_p\otimes V^{\otimes 3})\,$).
The subspace of invariant tensors in $V^*_p\otimes V^{\otimes 3}\,$
is two-dimensional.
We shall choose a basis $I_0\,,\ I_1\,$ in
it such that
\begin{equation}
\lb{2.15}
\Omega_{23}\, I_0 = 0\,,\quad I_1 = (P_{12} -1)\, I_0\
(\,\Rightarrow\ \Omega_{12}\, I_1 = 0\, )\,.
\end{equation}
Setting then
\begin{equation}
\lb{2.16}
F (\i )\, =\, (1-\i )\, f^0 (\i )\, I_0\, +\, \i\, f^1 (\i )\, I_1\,,
\end{equation}
we reduce the KZ equation to a system that does not depend on $N$:
\begin{equation}
\lb{2.17}
h (1-\i ) \frac{df^0}{d\i} = (h-2) f^0 + f^1\,,\quad h\,\i\,
\frac{df^1}{d\i} = (2-h) f^1 - f^0\,;
\end{equation}
it implies a hypergeometric equation for each $f^\ell\,:$
\begin{equation}
\lb{2.18}
\i (1-\i ) \frac{d^2 f^\ell}{d\i^2} + \left( 1+\ell -\frac{2}{h}-\left( 3-
\frac{4}{h}\right)\,\i\right) \frac{df^\ell}{d\i} =
\left(1-\frac{1}{h}\right)\left(1-\frac{3}{h}\right) f^\ell\,,\ \ell
=0,1\,.
\end{equation}

\section{Dynamical $R$-matrix exchange relations\\ 
among CVO}
\setcounter{equation}{0}
\renewcommand{\theequation}{3.\arabic{equation}}

Exchange relations among CVO provide an important ingredient in the finite
data characterizing a rational conformal field theory. They determine the
spectrum of anomalous dimensions (in other words, they allow a computation
of 
the conformal weights up to additive integers); they restrict the fusion
rules and determine the crossing symmetry (or ``duality") properties of
conformal blocks; they allow us to read off the statistics of
superselection
sectors, which has, according to 
[21 -- 23], an intrinsic
meaning in algebraic quantum field theory.

In order to derive the exchange properties of two ${\widehat{su}}
(N)\,$ step operators we shall consider the slightly more general
matrix element
\begin{equation}
\lb{3.1}
\Psi (p'' , p' ; z_1 , z_2 , z_3 ) = {\cal h} p'' |
\v (z_1 )\otimes \v (z_2 )\otimes \phi_{p'} (z_3 ) | 0{\cal i} =
D_{p'' p'} (z_{ab})\, F(\i )\,.
\end{equation}
Here $p'\,$ and $p''\,$ are the (shifted) weights of $SU(N)\,$ IR
such that the dimension of the space ${\cal I}_{p'' p'} = {\rm
Inv}\, (V^*_{p''}\otimes V^{\otimes 2}\otimes V_{p'} ) \ (\ni F (\i
))\,$ is maximal, ${\rm dim}\, {\cal I}_{p'' p'} = 2\,,$ and
\begin{equation}
\lb{3.2}
D_{p'' p'} (z_{ab}) =
z_{13}^{\D (p'')-\D (p')-2\D}\,
\i^{\frac{\D (p'' ) - \D (p' )}{2} - \frac{\fp}{2h} + \frac{2-N^2}{2Nh}}
(1-\i )^{-\frac{N+1}{Nh}}\,.
\end{equation}
In (\ref{3.2}) $\fp = p'_{ij}\ (\ge 2 )\,$ for $\v (z_a)\,,\
a=1,2\,,$ identified with the CVO $\v_i (z_1 )\,$ and
$\v_j (z_2 )\,,\ i<j\,,$ respectively (for a
synopsis of background material concerning $su(n)\,$ weights and the
corresponding conformal dimensions, see Appendix B). We
proceed with a summary of relevant results of \cite{HST00}.

The KZ equation for $\Psi\,$ again reduces to the form
(\ref{2.13}), only the expression (\ref{2.14}) and the relation
(\ref{A.9}) for $\Omega_{23}\,$ assume a more general form:
\begin{equation}
\lb{3.3}
\Omega_{23}= \frac{2-N}{2}
+h\,\frac{\Delta (p'') - \Delta (p')}{2} - C_{12} - C_{13}\,,\quad
\Omega_{23}^2 = \fp\,\Omega_{23}\ .
\end{equation}
(We recover (\ref{2.14}) and (\ref{A.9}) for ${p'}_{12} = 2\ 
(\, =\fp\, )\,,\ {p'}_{i\, i+1}=1\,$ for $2\le i \le N-1\,,$ in
which case $\Delta (p' )\,,\ \Delta (p'' ) = \Delta_{\phi}\,$
-- see (\ref{A.2}); another simple special case is $N=2\,,$ in
which $p' = p''\,.$) The relations (\ref{2.15}) for the basis
$\{ I_0 , I_1 \}\,$ of $SU(N)\,$ invariants remain unchanged while
the hypergeometric system (\ref{2.17}) assumes the form
\begin{equation}
\lb{3.4}
h (1-\i )\frac{d f^0}{d \i} = (h-2) f^0 + (\fp -1) f^1\,,
\quad h \i \frac{d f^1}{d \i} = (\fp - h) f^1 - f^0\,.
\end{equation}
A standard basis of (two) solutions is obtained by singling out the
possible analytic behaviour of the invariant amplitude $F (\i )\,$
(\ref{2.16}) for $\i \to 0\,.$ This gives the so-called {\em
$s$-channel basis} corresponding, in physical terms, to the {\em
operator product expansion} of $\v_j (z_2)\phi_{p'}(z_3) |0{\cal
i}\,$ (or of ${\cal h} 0 | \phi^*_{p''}(z_0)\v_i (z_1)\,$ -- see 
Appendix A). In the case at hand, these two solutions, 
$(f^\ell (\i ))_\l = s^\ell_\l (\i )\,,\ \l =0,1\,,$
are characterized by the property that $s^0_0 (\i )\,$
and $\i^{-\frac{\fp}{h}} s^0_1 (\i )\,$ are analytic and non-zero
at $\i = 0$:
\begin{equation}
\lb{3.5}
s^0_0 (0) = K_0\,,\quad s^0_1 (\i ) = K_1\,\i^{\frac{\fp}{h}} (1+ O(\i ))
\,,\quad K_\l \ne 0\,, \quad \l = 0, 1\,. 
\end{equation}
They are expressed in terms of hypergeometric functions:
\ba
&&s_0^0 (\i ) = K_0 \,
F\left( 1-\frac{1}{h},1-\frac{\fp +1}{h};1-\frac{\fp}{h};\i \right)\,,
\nonumber\\
&&s^1_0 (\i ) = K_0\, \frac{1}{\fp -h}\,
F\left( 1-\frac{1}{h},1-\frac{\fp +1}{h};2-\frac{\fp}{h};\i \right)\,,
\lb{3.6}
\\
&&s_1^0 (\i ) = K_1\, 
\i^{\frac{\fp}{h}}\, 
F\left( 1-\frac{1}{h},1+\frac{\fp -1}{h};1+\frac{\fp}{h};\i\right)\,,
\nonumber\\
&&s^1_1 (\i ) = K_1\, \frac{\fp}{\fp -1}\,
\i^{\frac{\fp}{h} -1} \,
F\left( -\frac{1}{h},\frac{\fp -1}{h};\frac{\fp}{h};\i \right)\,.
\lb{3.7}
\ea
We shall now compute the monodromy representation of the braid
group generator $B (\, B_1\, )\,$ corresponding to the exchange of
two ``identical particles" $1\,$ and $2\,.$ Note first that $\v\,$
(\ref{3.1}) is single-valued analytic in the neighbourhood of the
real configuration of points $\{\, z_1 > z_2 > z_3 > -z_2\,\}\,.$
We then choose any path in the homotopy class of
\begin{equation}
\lb{3.8}
\stackrel{\curvearrowright}{1\, 2}\,:\quad z_{1,2}(t) =
\frac{z_1+z_2}{2}\,\pm\frac{1}{2}\, z_{12}\, e^{-i\pi t}\,,\ \ z_3(t) =
z_3\,,\quad 0\le t\le 1\,,
\end{equation}
which thus exchanges $z_1\,$ and $z_2\,$ in a clockwise direction
and perform an analytic continuation of $\Psi\,$ along it,
followed by a permutation of the $SU(N)\,$ indices $A_1\,$ and
$A_2\,.$ This gives
\begin{eqnarray}
&&z_{12}\,\rightarrow\, e^{-i\pi} z_{12}\,,\quad
z_{13}\,\leftrightarrow\, z_{13}\,,\quad 1-\i \,\rightarrow\,
e^{-i\pi}\frac{1-\i}{\i}\quad
\left( \,\i \rightarrow \frac{1}{\i}\,\right) \,,\qquad \\
&&D_{p'' p'} (z_{ab})\ \rightarrow\
\bq^{\frac{N+1}{N}}\,\i^{\frac{\fp +1}{h}}\, D_{p'' p'} (z_{ab})
\quad {\rm for}\quad q^{\frac{1}{N}} = e^{-\frac{i\pi}{Nh}}  \,,\\
&&(1-\i )\, I_0\ \rightarrow\, -\frac{1-\i }{\i}\,
(I_0 + I_1 )\,,\quad \i\, I_1\,\rightarrow\, -\frac{1}{\i }\, I_1\ .
\end{eqnarray}
Using known transformation properties (the ``Kummer identities")
for hypergeometric functions (or rederiving them from their
integral representations -- see \cite{STH92}), we end up with the braid
relation
\begin{equation}
\stackrel{\curvearrowright}{1\, 2}\,:\quad
D\, s^{\ell}_\l (\i )\ \stackrel{\curvearrowright}\to\
D\, s^{\ell}_{{\l}'}
(\i )\, B^{{\l}'}_\l\,,\ \quad
B = \bq^{\frac{1}{N}}\,
\left(
\begin{array}{ccc}\frac{q^\fp}{[\fp ]}&Kb_{\fp}\\
K^{-1}{b_{-\fp}}& -\frac{\bq^{\fp}}{[{\fp}]}\end{array}
\right)\,,  
\label{3.12}
\end{equation}
where $K=\frac{K_1}{K_0}\,$ and
\begin{equation}
\lb{3.13}
[\fp ] := \frac{q^{\fp}-\bq^{\fp}}{q-\bq}\,,\
b_\fp = \frac{\G (1+\frac{\fp}{h}) \G (\frac{\fp}{h}) }
{\G (1+\frac{\fp -1}{h}) \G (\frac{\fp +1}{h}) }\
\left(\,\Rightarrow b_\fp b_{-\fp} =
\frac{[\fp +1][\fp -1]}{[\fp]^2}\,\right)\,.
\end{equation}
The eigenvalues of $B\,$ do not depend on the choice of the relative
normalization, $K = K(\fp )\,,$ and turn out to be also $\fp$-independent. 
Indeed, they are expressed in terms of the dimension 
$\D = \D (\L_1 )\,$ (\ref{A.2}) of a step operator and by the dimensions
$\D_s = \D (2\L_1 )\,$ and $\D_a = \D (\L_2 )\,$ of the symmetric and
skew-symmetric tensor products of two fundamental (``quark")
representations
of $SU(N)$:
\ba
\lb{3.13A}
&&{\rm spec}\, (B) = (e^{i\pi (2\D -\D_s )}\,,\; - e^{i\pi (2\D -\D_a )}\, ) =
(q^{\frac{N-1}{N}}\,,\; - \bq^{\frac{N+1}{N}}\, )\,,\nonumber\\ 
&&\D_s = \frac{(N-1)(N+2)}{Nh}\,,\quad 
\D_a = \frac{(N+1)(N-2)}{Nh}\,.
\ea
It is
remarkable that for $K(\fp )\, K(-\fp )\, = \, 1\,,$ in particular, for  
\begin{equation}
\lb{3.14}
K= \frac{\G (\frac{1+\fp}{h}) \G (-\frac{\fp}{h}) }
{\G (\frac{1-\fp}{h}) \G (\frac{\fp}{h}) }\rho (\fp )\,,\quad
\rho (\fp ) \rho (-\fp ) = 1\  (\, = K(\fp )K(-\fp )\,)\,,
\end{equation}
(\ref{3.12}) agrees with the (dynamical) $R$-matrix exchange
relations
\begin{equation}
\lb{3.15}
\stackrel{\curvearrowright}
{\v^B_i (z_2)\, \v^A_j (z_1)}\, =\, \v^A_s (z_1)\, \v^B_t
(z_2)
\, \R (\fp )^{st}_{ij}
\end{equation}
linked in \cite{FHIOPT} with the properties of an intertwining
quantum matrix algebra generated by an $N\times N\,$ matrix
$(a^i_\a )\,$ with non-commuting entries and by $N\,$ commuting
unitary operators $q^{p_i}\,\ ( \prod_{i=1}^N\, q^{p_i} = 1\,)\,,$
such that
\begin{equation}
\lb{3.16}
q^{p_i}\, a^j_\a\,=\, a^j_\a\, q^{p_i + \d^j_i -\frac{1}{N}}\,,\quad
\Rp \, a_1 \, a_2\, =\, a_1 \, a_2 \, \R\,,\quad
\v^A_\a (z) = \v^A_i (z)\, a^i_\a\,.
\end{equation}
Here $\R = ( \R^{\a_1\a_2}_{\b_1\b_2})\,$
and $\Rp = (\Rp^{i_1 i_2}_{j_1 j_2}\,$ are the $U_q(sl_N )\,$ and
the dynamical $R$-matrices, respectively, multiplied by a
permutation, $\R = R\, P\,$ and we are using Faddeev's concise
notation for tensor products. $\Rp\,$ obeys the Gervais--Neveu
\cite{GN84} ``dynamical Yang--Baxter equation" whose general solution
satisfying ``the {\em ice condition}" (the condition that
$\Rp^{ij}_{kl}\,$ vanishes unless the {\em unordered} pairs
$(i,j)\,$ and $(k,l)\,$ coincide) was found by A. Isaev
\cite{Isa96}. Its $2\times 2\,$ block
\begin{equation}
\left(\begin{array}{ccc}
\R (p)^{ij}_{ij}& \R (p)^{ij}_{ji}\\
\R (p)^{ji}_{ij}& \R (p)^{ji}_{ji}
\end{array}\right) = \bq^{\frac{1}{N}}
\left(\begin{array}{ccc}
\frac{q^\fp}{[\fp ]}&\frac{[\fp -1]}{[\fp ]}\rho(\fp )
\\ \frac{[\fp +1]}{[\fp ]}{\rho (-\fp)}&-\frac{\bq^\fp}{[\fp ]}
\end{array}\right)\quad (\, \rho (\fp )\rho (-\fp ) = 1\, )
\label{3.17}
\end{equation}
indeed coincides with $B$, for $K\,$ given by (\ref{3.14}).

The monodromy representations of the braid group in the space of
solutions of the KZ equation was first studied systematically in
\cite{TK87}. The Drinfeld--Kohno theorem \cite{Ko87,Dr89} (see also
\cite{Ka95}, 
Ch. 19) says, essentially, that for generic $q\,$ this monodromy
representation is always given by (a finite-dimensional
representation of) Drinfeld's universal $R$-matrix. In the
physically interesting case of $q$, an (even) root of unity ($\, q^h
= -1\,$), the situation is more complicated. A problem already
appears in Eq. (\ref{3.17}): for $\fp = h\,,\ [h]=0\,$ and the right-hand 
side of (\ref{3.17}) makes no sense. In fact, the
representation of the braid group is not unitarizable for such
values of $\fp\,.$ The corresponding ``unphysical" solutions of the
KZ equation cannot, on the other hand, be thrown away by decree;
otherwise, the chiral field algebra will not be closed under
multiplication.

It turns out that a monodromy representation of the braid group
can, in fact, be defined on the entire space of solutions of the
KZ equation. It is, in general, indecomposable. The above
$s$-channel basis, however, does not extend to $\fp = h\,$ (as
is manifest in Eq. (\ref{3.6})).

\section{Regular basis of solutions of the KZ equation 
and Schwarz finite monodromy problem}
\setcounter{equation}{0}
\renewcommand{\theequation}{4.\arabic{equation}}

It follows from (\ref{3.15}) and (\ref{3.16}) that the chiral
fields (unlike the CVO $\v^A_j\,$) satisfy $\fp$-independent (and
hence, non-singular) exchange relations:
\begin{eqnarray}
&&\stackrel{\curvearrowright}
{\v^B_\a (z_2)\, \v^A_\b (z_1)}\, =\, \v^A_{\rho} (z_1)\, \v^B_{\s}
(z_2)
\, \R^{\rho\s}_{\a\b}\,,
\lb{4.1}\\
&&\R = \bq^{\frac{1}{N}} (q\id - A )\,,\qquad A^{\rho\s}_{\a\b} =
q^{\epsilon_{\s\rho}} \d^\rho_\s\,\d^\s_\b\, -\,
\d^\rho_\b\,\d^\s_\a\,,\nonumber\\
&&\epsilon_{\s\rho} =
\left\{
\begin{array}{ccc}
1\,, & \s > \rho \\
0\,, & \s =\rho \\
-1\,, & \s < \rho
\end{array}
\right .
\,,\qquad A^2\, =\, [2]\, A\,,\quad [2]=q+\bq\,.
\lb{4.2}
\end{eqnarray}
The singularity in the conformal block (\ref{3.6}) for $\fp = h\,$
is thus a consequence of the introduction of CVO which pretend to
diagonalize the (in general, non-diagonalizable) monodromy matrix
$M\,$ defined by $\v\, (z\,e^{2\pi i})\, =\, \v\, M\,.$ A regular
basis of conformal blocks is linked to a regular basis in $U_q
(sl_N )\,$ invariant tensors (with respect to the indices $\a ,
\b ,\dots\,$). Such a basis has been introduced for $N=2\,$ in
\cite{FST91} and recently generalized to four-point blocks involving a
pair of $U_q (sl_N )\,$ step operators \cite{HST00}. Its counterpart in
the space of conformal blocks of the $SU(N)\,$ WZNW model was
written down in \cite{STH92} (for $N=2\,$) and in \cite{HST00} for
arbitrary
$N\,.$ We shall display here a regular basis of four-point
conformal blocks $f^\ell_\l\,,$ only mentioning in conclusion some
properties of their quantum group counterparts ${\cal I}^\l\,.$

Writing the M\"obius-invariant amplitude (\ref{2.16}) in the form
\begin{equation}
\lb{4.3}
F(\i ) = F_0 (\i )\,{\cal I}^0 + F_1 (\i )\,{\cal I}^1\,,\quad F_\l
(\i ) = (1-\i ) f^0_\l (\i ) I_0 + \i f^1_\l (\i ) I_1\,,
\end{equation}
where ${\cal I}^\l\,$ are $U_q(sl_N )\,$ invariant tensors to be
specified below, we define the regular basis by
\begin{eqnarray}
\lb{contour}
&& B \left(\frac{\fp -1}{h} , \frac{2}{h} \right)\, f^{\ell}_0 (\i ) =
\,\int_{\i}^1 t^{\frac{\fp -1}{h} -\ell}
(1-t)^{\frac{1}{h}-1+\ell}(t-\i )^{\frac{1}{h}-1} dt =
\lb{4.4}
\\ &&=
B \left(\frac{1}{h} , \ell + \frac{1}{h} \right) (1-\i
)^{\frac{2}{h}-1+\ell}
F \left(\ell - \frac{\fp -1}{h} ,
\ell + \frac{1}{h} ; \ell + \frac{2}{h} ; 1-\i \right)\,,\nonumber\\
&&B \left(\frac{\fp -1}{h} , \frac{2}{h} \right)\, f^{\ell}_1 (\i ) =
\int^{\i}_0 t^{\frac{\fp -1}{h} -\ell}
(1-t)^{\frac{1}{h}-1+\ell}(\i -t)^{\frac{1}{h}-1} dt =
\lb{4.5}
\\ 
&&= B\left(\frac{1}{h} , 1-\ell + \frac{\fp -1}{h} \right)
\i^{\frac{\fp}{h}-\ell}
F \left( 1-\ell - \frac{1}{h} ,1-\ell + \frac{\fp -1}{h} ; 1-\ell +
\frac{\fp}{h} ; \i \right)\nonumber
\end{eqnarray}
$\left( B(\mu , \nu ) = \frac{\G (\mu ) \G (\nu )}{\G (\mu +\nu
)}\right)\,.$ A direct computation using the integral
representations for $f^\ell_\l (\i )\,$ yields the following form
for the braid matrix $B_1\,,$ exchanging the arguments $1\,$ and
$2\,$ (the counterpart of $B\,$ (\ref{3.12}) in the regular basis,
$B\,$ and $B_1\,$ having the same eigenvalues):
\ba
\lb{4.6}
&&B_1 = \bq^{\frac{1}{N}}\left(\begin{array}{ccc}q&1\\ 
0&{-\bq}
\end{array}\right)\,,\nonumber\\ 
&&{\rm det}\, (q^{\frac{1}{N}} B_1 ) = -1 = {\rm
det}\,(q^{\frac{1}{N}} B )\,,
\quad  {\rm tr}\, (q^{\frac{1}{N}} B_1 ) = q-\bq =
{\rm tr}\,(q^{\frac{1}{N}} B )\,.
\ea
$B_1\,$ and $B\,$ are thus related by a similarity transformation
whenever both make sense:
\begin{equation}
\lb{4.7}
B_1 = S B S^{-1}\,,\ S = \left(\begin{array}{ccc}1&0 \\
-\frac{[\fp -1]}{[\fp ]}&\rho (\fp )\frac{[\fp -1]}{[\fp ]}
\end{array}\right)\,,\  S^{-1} =\left(\begin{array}{ccc}1&0\\ 
\rho (-\fp )&\rho (-\fp )\frac{[\fp ]}{[\fp -1]} \end{array}\right)\,.
\end{equation}
Similarly, the exchange matrix $B_2\,$ (corresponding to the
braiding $\stackrel{\curvearrowright}{2\, 3}\,$) is given by
\begin{equation}
\lb{4.8}
B_2 = S q^{\frac{1-\fp}{N}}\left(\begin{array}{ccc}{-\bq}&0\\ 0& q^{\fp
-1}\end{array}\right) S^{-1} = q^{\frac{1-\fp}{N}}
\left(\begin{array}{ccc}{-\bq}&0\\ 
\frac{q^{2(\fp -1)}-1}{q^{\fp}-1}
&q^{\fp -1}\end{array}
\right)\,.
\end{equation}
The eigenvalues of $B_2\,,$ like those of $B_1\,,$ cf. (\ref{3.13A}), are
expressed in terms of conformal dimensions. 
Setting $i=1\,,\ j=2\,,$ assuming that $p'\,$ is the symmetric tensor
representation (\ref{B.3}) and 
\be
\lb{intermezzo1}
p_s = (\fp +1,1,\dots ,1\,)\,,\qquad p_a = (\fp -1,2,\dots ,1\,)
\ee 
(in the notation $p = (p_{12},p_{23},\dots ,p_{N-1\, N}\,)$), we find 
\be
\lb{intermezzo2}
e^{i\pi (\D +\D (p) -\D (p_a ))} = q^{\frac{1-\fp -N}{N}}\,,\qquad
e^{i\pi (\D +\D (p) -\D (p_s ))} = q^{(\fp -1)\frac{N-1}{N}}\,
\ee
for 
\be
\lb{intermezzo3}
\D (p_a ) = \frac{(N-1)(\fp -2)^2 + (N-2)(N+1)\fp}{2Nh}\,,\quad
\D (p_s ) = \frac{(N-1)\fp (\fp +N)}{2Nh}\,.
\ee
We observe that, unlike $B$, the matrices $B_1\,$ and $B_2\,$
are defined for $0<\fp <2h\,.$ The singularity in $S\,$ (\ref{4.7}), as
well as the non-existence of the $s$-channel basis for $\fp = h\,,$
is due to the simple fact that the matrix $B_2\,$ (\ref{4.8}) is
non-diagonalizable in this case (while the $s$-channel basis could
be defined as ``the basis in which $B_2\,$ is diagonal"). Note that,
for $\fp = 2\,,\  B_2\,$ becomes similar to $B_1$:
\begin{equation}
B_2 = \bq^{\frac{1}{N}}\left(
\begin{array}{ccc}{-\bq}&0\\
1&q\end{array}
\right) = \s_1 B_1\s_1\,,\quad 
\s_1 =\left(
\begin{array}{ccc}0&1\\ 
1&0\end{array}
\right)\  \
(\,{\rm for}\ \fp=2\,)
\lb{4.9}
\end{equation}
and $B_1\,$ and $B_2\,$ generate a representation of the braid
group ${\cal B}_3\,$ with central element $c^3 = (B_1 B_2 )^3 =
\bq^{\frac{6}{N}}\,\id\,.$

Whenever the $s$-channel basis (\ref{3.6}), (\ref{3.7}) exists, it is
related to
the regular basis (\ref{4.3})--(\ref{4.5}) by
\begin{equation}
\lb{4.10}
F_0 (\i )\,{\cal I}^0 + F_1 (\i )\,{\cal I}^1 = s_0 (\i )\,{\cal
S}^0 + s_1 (\i )\,{\cal S}^1 \,.
\end{equation}
Here ${\cal I}^0\,$ and ${\cal S}^0\,$ are equal and can be expressed as 
a matrix element of a product of $a^i_\a\,$ satisfying (\ref{3.17}):
\begin{equation}
\lb{4.11}
( {\cal S}^0 = {\cal I}^0 = ) \,{\cal h} p'' | a^i_{\a_1}
a^j_{\a_2} | p'{\cal i}\quad {\rm for}\ i<j\,;
\qquad {\cal S}^1= {\cal h}
p'' | a^j_{\a_1} a^i_{\a_2} | p'{\cal i}\,.
\end{equation}
(If $p'\,$ is the symmetric tensor representation, see (\ref{B.3}), 
then we choose
$i = 1\,,\, j=2\,.$) The invariant tensor ${\cal I}^1\,,$ on the other
hand, is related to ${\cal I}^0\,$ by
\begin{equation}
\lb{4.12}
{\cal I}^1_{\dots\a_1\a_2\dots} = \, -\,{\cal
I}^0_{\dots\s_1\s_2\dots}\, A^{\s_1\s_2}_{\a_1\a_2}\,,
\end{equation}
where $A\,$ is the quantum antisymmetrizer defined in (\ref{4.2}).
The exchange relations (\ref{3.16}) with the dynamical $R$-matrix
(\ref{3.17}) then allow us to relate ${\cal S}^\l\,$ with ${\cal
I}^\l\,$ and conversely:
\begin{equation}
\lb{4.13}
\rho (\fp )\, {\cal S}^1_{\a\b}\, =\, {\cal I}^0_{\a\b}\, +\,
\frac{[\fp ]}{[\fp -1]}\,{\cal I}^1_{\a\b}\,,\quad
{\cal I}^1_{\a\b} \, =\, \frac{[\fp -1]}{[\fp ]}\,
\left(\,
\rho (\fp )\, {\cal S}^1_{\a\b}\,-\,{\cal S}^0_{\a\b}
\right)\,.
\end{equation}
For $\fp\  (= {p'}_{ij} ) = h\,,\ {\cal S}^0\,$ and ${\cal S}^1\,$ are
proportional, ${\cal S}^0=\rho (h) {\cal S}^1\,,$ so that they do
not form a basis; ${\cal I}^1\,,$ on the other hand, is defined
unambiguously by (\ref{4.12}) and is linearly independent of ${\cal
I}^0\,.$

The above regular basis also has a remarkable number theoretic
property: the matrix elements of $q^{\frac{1}{N}} B_1\,$ (and of
$q^{\frac{\fp -1}{N}} B_2\,$) belong to the {\em cyclotomic field}
${\Q} (q)\,$ of polynomials in $q\,$ with rational coefficients for
$q^h = -1\,.$ This fact has been used in
\cite{ST95} to classify all cases in which the monodromy representation
of the braid group 
${\cal B}_4\,,$ for $N=2\,,$ is
a finite matrix group or, equivalently, the cases in which the KZ
equation has an algebraic solution (a classical problem solved for
the hypergeometric equation by H.A. Schwarz in the 1870's). As pointed
out in \cite{HST00}, this result readily extends to higher $N\,$ in the 
case of three step operators (for ${\cal B}_3\,$). 
The argument uses one of the oldest and most beautiful concepts
in group theory, the Galois group, so that it deserves to be summarized.

The space of $U_q(sl_N )\,$ invariants admits a braid invariant
hermitean form $(\,,\, )\,.$ In the regular basis, $Q^{\l\mu}
\equiv ({\cal I}^\l\,, {\cal I}^\mu )\,$ belong to the real
subfield ${\Q} ([2]) = {\Q} (q+\bq )\,$ of ${\Q}
(q)\,.$ The special case of $N=2\,$ is worked out in Appendix C. In
that case the resulting hermitean form $Q\,$ is positive-semidefinite 
for $q= e^{\pm i\frac{\pi}{h}}\,$ and has a kernel of
dimension $2\fp - h\,$ for $2\,\fp >h\,.$ For the case $\fp = 2\,$ of
interest this kernel is only nontrivial at level $1\,,$ for
$h=3\,,$ when it is one-dimensional.

We define a {\em primitive root} of the equation $q^h =-1\,$ as 
any zero of the irreducible element $P_h (q)\,$ of the ring of
polynomials with integer coefficients satisfying $P_h (e^{\pm
i\frac{\pi}{h}}) = 0\,.$ (There is a unique such irreducible
polynomial with coefficient to the highest power of $q\,,$ equal to
$1\,$.) The {\em Galois group} ${\rm Gal}_h\,$ for $P_h\,,$ the
group that permutes its roots, consists of all substitutions of the
form
\begin{equation}
\lb{4.14}
{\rm Gal}_h = \{ \, q\, \to\, q^\ell\,,\ 0<\ell <2h\,,\ 
(\ell\,,\, 2h) = 1\,\}
\end{equation}
(in the last condition in the definition we use the familiar notation
$(\ell\,,\, m)\,$ for the greatest common divisor of $\ell\,$ and $m\,$).

A hermitean form with entries in a
cyclotomic field ${\Q} (q)\,$ is called {\em totally positive}
if all its Galois transforms are positive. Our analysis is based on
the following theorem. The total positivity of a ${\cal
B}_n$-invariant form $Q\,$ is sufficient (and, if the invariant
form is unique, also necessary) for the monodromy representation of
${\cal B}_n\,$ in $q$-Inv$(V^{\otimes n})\, / {\rm Ker}\, Q\,$ to be
a finite matrix group. For $N = 2\,$ and $h>3\,$ we find \cite{ST95}
that the total positivity of $Q\,$ is equivalent to the total
positivity of the quantum dimension $[3] = \frac{q^3 -\bq^3}{q-\bq}
\equiv q^2 + 1 + \bq^2\,$ encountered in the tensor product
expansion of the tensor square of the two-dimensional
representation: $[2]^2 = 1+[3]\,.$ This amounts to finding the
values of $h\ge 4\,$ such that
\begin{equation}
\lb{4.15}
1\, +\,\cos\, \frac{2\pi\ell}{h}\, > \,0\quad{\rm for}\quad (\ell ,
2h ) = 1\,.
\end{equation}
The only solutions are $h = 4, 6 , 10\,.$ If we add to these the
case $h=3\,,$ in which the commutator subgroup of ${\cal B}_4\,$ is
trivial $(\, B_0 B_1 B_0^{-1} B_1^{-1} = 1 = B_1 B_2 B_1^{-1}
B_2^{-1} = \dots\,)\,,$ we see that the four cases of ``finite
monodromy" correspond to the four integral quadratic algebras of
dimension $h-2 = 1,2,4,8\,.$


\section*{Acknowledgements}

I.T. thanks John Roberts for an enlightening correspondence 
concerning the scope of the main theorem of \cite{DR90} and George
Pogosyan for his invitation to the Group Theoretical Methods Colloquium.
The authors thank the Theory Division of CERN for hospitality
and acknowledge partial support from the Bulgarian
National Council for Scientific Research under contract F-828.

\section*{Appendix A. Reduction of the KZ equation for $SU(N)\,$
step operators to an $N$-independent system of
hypergeometric equations}
\setcounter{equation}{0}
\def\theequation{A.\arabic{equation}}

The ``wave function" $\Psi (p ; z_1 , z_2 , z_3 )\,$ can be viewed
as a $z_0 \to \infty\,$ limit of a M\"obius and $SU(N)$-invariant 
four-point function
\begin{equation}
\lb{A.1}
w (z_0 ; z_1 , z_2 , z_3 ) = {\cal h} 0 | \phi^* (z_0)\otimes \v (z_1)
\otimes \v (z_2 )\otimes \v (z_3) | 0 {\cal i}\,,\quad \phi (0) |
0{\cal i}\, \sim \,
\put(0,2){\framebox(5,5)}
\put(0,-3){\framebox(5,5)}
\put(5,2){\framebox(5,5)}\quad\,.
\end{equation}
The step operator $\v\,$ and the field $\phi\,$ have $su(N)\,$
weights $\Lambda_1\,$ and $\Lambda_1 + \Lambda_2\ \,
(\Lambda_j\,,\ j=1,\dots , N-1\,$ being the fundamental $su(N)\,$
weights). Their conformal dimensions are
\ba
\lb{A.2}
&&\Delta = \Delta (\Lambda_1 ) = \frac{1}{2h} C_2 (\Lambda_1 ) =
\frac{N^2 -1}{2hN}\,,\nonumber\\ 
&&\Delta_\phi = \Delta (\Lambda_1+\Lambda_2 ) 
= \frac{1}{2h} C_2 (\Lambda_1+\Lambda_2 ) = \frac{3}{2hN} (N^2 - 3)\,,
\ea
where $C_2 (\Lambda )\,$ stands for the eigenvalue of the second-order 
Casimir invariant (normalized in such a way that for the
adjoint representation $C_2 (\Lambda_1+\Lambda_{N-1} ) = 2N\,$).
M\"obius (i.e. $SL(2)\,$) invariance implies that we can write
$w\,$ in the form
\begin{equation}
\lb{A.3}
w (z_0 ; z_1 , z_2 , z_3 ) = D_N (z_{ab})\,F(\i )\,,\quad
\i = \frac{z_{01}z_{23}}{z_{02}z_{13}}\,.
\end{equation}
The prefactor $D_N\,$ is a product of powers of the coordinate differences
$z_{ab}\,$ determined from infinitesimal M\"obius invariance
\begin{equation}
\lb{A.4}
\left( z_0^\nu (z_0\frac{\partial}{\partial z_0} + (\nu
+1)\Delta_\phi ) + \sum_{c=1}^3 z_c^\nu 
(z_c\frac{\partial}{\partial z_c} + (\nu
+1)\Delta ) \right) D_N (z_{ab}) = 0
\end{equation}
for $\nu=0,\pm 1\,,$
up to powers of $\i\,$ and $(1-\i )\,,$ which are fixed by
requiring that there exists a solution $F(\i )\,$ of the
resulting ordinary differential equation that takes finite
non-zero values for both $\i = 0\,$ and $\i = 1$:
\begin{equation}
\lb{A.5}
D_{N}(z_{ab}) = \left(\frac{z_{13}^{2N+4}}{z_{03}^{3N+5} z_{12}}
 \right)^{\frac{N-2}{2Nh}}
\frac{(1-\i )^{\frac{(N-4)(N+1)}{2Nh}}}{z_{01}^\frac{N^2+N-3}{Nh}
z_{23}^\frac{N+1}{Nh} } = \left(\frac{z_{13}^{-3} (\i (1-\i
))^{-N-1}}{z_{02}^{N^2 -1} (z_{01} z_{03})^{N^4-4}} \right)^\frac{1}{Nh}\,.
\end{equation}
Comparing the last expression with (\ref{2.12}) we find the relation
\ba
&&\Psi_p (z_1 , z_2 , z_3 ) = \lim_{z_0\to\infty}\{
z_0^{2\Delta_\phi} w (z_0 ; z_1 , z_2 , z_3 )\} =\nonumber\\ 
&&= \,\left( z_{02}^{N^2-1} (z_{01}z_{03})^{N^2-4}\right)^\frac{1}{Nh}
w (z_0 ; z_1 , z_2 , z_3 )\,.
\lb{A.6}
\ea
Applying to the $4$-point function $w\,$ (\ref{A.3}) the
covariant derivative $h\nabla_1\,$ (\ref{2.4}),
\begin{equation}
\lb{A.7}
h\nabla_1 = h\frac{\partial}{\partial z_1} +
\frac{C_{01}}{z_{01}} - \frac{C_{12}}{z_{12}} - \frac{C_{13}}{z_{13}}\,,
\end{equation}
and using the $SU(N)\,$ invariance condition
\begin{equation}
\lb{A.8}
(C_{01} + C_{12} + C_{13} + C_2 (\Lambda_1 ) )\, 
w (z_0 ; z_1 , z_2 , z_3 )\, = 0\,,
\end{equation}
we end up with (\ref{2.13}) for $\Omega_{ab}\,$ given by
(\ref{2.14}). The operators $\Omega_{12}\,$ and $\Omega_{23}\,$
have an algebraic characterization of the Temperley--Lieb type (see
\cite{HST00}, Eq. (2.14)):
\begin{equation}
\lb{A.9}
\Omega_{12}\Omega_{23}\Omega_{12} = \Omega_{12}\,,\quad 
\Omega_{23}\Omega_{12}\Omega_{23} = \Omega_{23}\,,\quad
\Omega_{ab}^2 = 2\,\Omega_{ab}\,.
\end{equation}
In particular, each $\Omega_{ab}\,$ has eigenvalues $0\,$ and
$2\,.$ If we regard $\phi^*\,$ as a mixed tensor of $2N-3\,$ indices,
$\phi^* = \{ (\phi^* )^{B_1 \dots B_{N-1}C_1\dots C_{N-2}} \}\,,$
then the $SU(N)\,$ invariant tensors $I_0\,$ and $I_1\,$ of Eq.
(\ref{2.15}) can be presented in the form
\begin{equation}
\lb{A.10}
I_0 = \left(\epsilon^{B_1\dots B_{N-1} A_1} \epsilon^{C_1\dots
C_{N-2}A_2 A_3 } \right)\,, \quad I_1 = (P_{12}-1)\,I_0\,,
\end{equation}
where $\epsilon\,$ is the totally antisymmetric Levi-Civita
tensor, and $P_{12}\,$ permutes the indices $A_1\,$ and $A_2\,.$ In
this basis the operators $\Omega_{12}\,$ and $\Omega_{23}\,$ have
the following matrix realization:
\begin{equation}
\lb{A.11}
\Omega_{12} = \left(\begin{array}{ccc}2&1\\ 0&0\end{array}
\right)\,,\quad
\Omega_{23} = \left(\begin{array}{ccc}0&0\\ 1&2\end{array} \right)\,,
\end{equation}
i.e. $\Omega_{12} I_0 = 2 I_0 + I_1\,,\ \Omega_{23} I_1 = I_0 + 2
I_1\,,$ etc. Remarkably, the relations (\ref{A.9}) and (\ref{A.11})
are independent of $N\,.$ Inserting (\ref{2.16}) into
(\ref{2.13}) and using (\ref{A.11}), we thus end up with the
$N$-independent system (\ref{2.17}) of a hypergeometric type.

\section*{Appendix B. Shifted $SU(N)\,$ weights and CVO.
Symmetric tensor representations}
\setcounter{equation}{0}
\def\theequation{B.\arabic{equation}}

If $\Lambda = \sum_{i=1}^{N-1}\, \l_i\Lambda_i\,,\ \l_i \in {\Z}_+\,,$
is an $su(N)\,$ {\em highest-weight} ($\l_i\,$ being the number
of columns of height $i\,$ in the associated  Young tableau), then
the corresponding {\em shifted weight} is written in terms of
barycentric coordinates $p = (p_1 ,\dots , p_N )\,$ as follows:
\begin{equation}
\lb{B.1}
p = \Lambda + \rho = \sum_{i=1}^{N-1} p_{i\, i+1}
\Lambda_i\,,\quad p_{ij} = p_i - p_j\,,\quad p_{i\, i+1} = \l_i +1\,,\quad
\sum_{i=1}^N p_i = 0
\end{equation}
($\rho = \sum_{i=1}^{N-1} \L_i\,$ is the half sum of the
positive roots.)
The conformal dimension of a ${\widehat{su}} (N)\,$ primary field
of weight $p\,$ is expressed in terms of the second-order Casimir
operator $C_2 (p)\,:$
\begin{equation}
\lb{B.2}
2h \Delta (p) = C_2 (p) = \frac{1}{N} 
\sum_{i<j}\left(p_{ij}^2 - (j-i)^2 \right) 
= \frac{1}{N} \sum_{i<j} p_{ij}^2 - \frac{N(N^2-1)}{12}\,.
\end{equation}
The CVO $\v_j (\, = \v_j^A (z)\, )\,$ is related to the $U_q
(sl_N )\,$ covariant field $\v_\a (\, = \v_\a^A (z)\, )\,$ by (\ref{3.16}).

In the example of a symmetric tensor representation $p'\,$ and
its counterpart $p''\,$ defined by the requirement ${\rm dim}\ 
{\cal I}_{p' p''} = 2\,$ we have
\begin{eqnarray}
\lb{B.3}
&&{p'}_{12}=p\,,\quad {p'}_{i\,i+1} = 1\,,\ 2\le i\le
N-1\,,\nonumber\\ 
&&C_2(p' )= \frac{N-1}{N} (p-1)(p+N-1)\,,\\
&&{p''}_{12} = p\,,\quad {p''}_{23}=2\,,\quad {p''}_{i\,i+1}=1\,,\
3\le i\le N-1\,,\nonumber\\ 
&&C_2 (p'' ) = (p+1)\,
\frac{N^2+(p-2)N-(p+1)}{N}\,.
\end{eqnarray}
The dimensions of these representations are expressed in terms of
binomial coefficients:
\begin{equation}
\lb{B.5}
d (p' ) = {{p+N-2}\choose{N-1}}\,,\quad
d (p'' ) = p \, {{p+N-1}\choose{p+1}}\,.
\end{equation}
In computing the prefactor (\ref{3.2}) one needs
\begin{eqnarray}
&&\Delta (p'' ) - \Delta (p' ) - 2 \Delta =\nonumber\\
\lb{B.6}
&&=\, \frac{(N-2)(N+p)}{Nh} - \frac{N^2-1}{Nh} =
\frac{(p-2)(N-2)-3}{Nh}\,,\\
\lb{B.7}
&&\frac{\Delta (p'' )- \Delta (p' )}{2} - \frac{p}{2h} - \frac{N^2-2}{2Nh}
= -\, \frac{N+p-1}{Nh}\,.
\end{eqnarray}

\section*{Appendix C. Basis of $U_{q}(sl_2)\,$ invariants in
$V^{\otimes 4}\,$ for $V = {\C}^2\,.$ Braid-invariant
hermitean form}
\setcounter{equation}{0}
\def\theequation{C.\arabic{equation}}

The basic $U_q(sl_N )\,$ invariant in $V^{\otimes N}\,$ for $V =
{\C}^N\,$ is the $q$-deformed Levi-Civita tensor
\begin{equation}
\lb{C.1}
{\cal E}_{\a_1\dots \a_N}= \bq^{ \frac{1}{2} {N\choose{2}}}
(-q^2 )^{\ell
{\scriptsize{\left(\begin{array}{ccc}N&\dots &1\\ \a_1
&\dots&\a_N\end{array}\right)}}
}\,,
\end{equation}
where $\ell\,$ is the length of the permutation
$\scriptsize{\left(\begin{array}{ccc}N&\dots &1\\ \a_1 &\dots
&\a_N\end{array}\right)}\,,$ i.e.
the minimal number of transpositions of neighbouring indices; in
particular, for $N=2\,$ 
\begin{equation}
\lb{C.2}
\left( {\cal E}_{\a_1\a_2}\right) = \left(\begin{array}{ccc}0&
-q^{\frac{1}{2}}\\
\bq^{\frac{1}{2}}&0\end{array}\right)\,,\quad {\rm i.e.}\ \  
{\cal E}_{21} = \bq^{\frac{1}{2}} \,,\ \ {\cal E}_{12} = - q^\frac{1}{2}\,.
\end{equation}
The regular basis of $U_q (sl_2 )\,$ invariants in $V^{\otimes 4}
= ({\C}^2)^{\otimes 4}\,$ is
\begin{equation}
\lb{C.3}
{{\cal I}^0}_{\a_1\a_2\a_3\a_4} = 
{\cal E}_{\a_1\a_2}\,{\cal E}_{\a_3\a_4}\,,
\qquad {{\cal I}^1}_{\a_1\a_2\a_3\a_4} = 
{\cal E}_{\a_1\a_4}\,{\cal E}_{\a_2\a_3}\,.
\end{equation}
Their inner products are given by traces:
\ba
\lb{C.4}
&&( {\cal I}^\l\,, {\cal I}^\mu ) = \sum_{\a_1\dots\a_4} 
{{\cal I}^\l}_{\a_1\a_2\a_3\a_4}\,{{\cal
I}^\mu}_{\a_1\a_2\a_3\a_4} \,,\nonumber\\
&&( {\cal I}^\l\,, {\cal I}^\l ) =
[2]^2\,,\ \ \l = 0,1\,,\quad  
( {\cal I}^0\,, {\cal I}^1 ) = - [2]\,.
\ea
To verify braid invariance, note that
\begin{eqnarray}
&&{B_1}^0_\l {\cal I}^\l = q^\frac{1}{2} {\cal I}^0 +
\bq^\frac{1}{2} {\cal I}^1\,,\quad 
{B_1}^1_\l {\cal I}^\l = 
- \bq^\frac{3}{2} {\cal I}^1\,, 
\nonumber\\
&&( {B_1}^0_\l {\cal I}^\l\,, {B_1}^0_\mu {\cal I}^\mu ) = 2[2]^2
- (q+\bq )[2] = [2]^2 = ( {\cal I}^0\,, {\cal I}^0 )\,,\ \, {\rm etc.}
\end{eqnarray}

\end{document}